\documentclass[,final]{aipproc}

\layoutstyle{6x9}

\usepackage{amsmath}
\usepackage{amsfonts}
\usepackage{amssymb}
\usepackage{graphicx}

\input{tcilatex}

\begin{document}

\title{Updating Probabilities with Data and Moments\thanks%
{Presented at the 27th International Workshop on Bayesian Inference and
Maximum Entropy Methods in Science and Engineering, Saratoga Springs, NY,
July 8-13, 2007.)
}}

\classification{}
\keywords{relative entropy, Bayes theorem, expectation value, moment}

\author{Adom Giffin\thanks{%
E-mail: physics101@gmail.com} \ and Ariel Caticha\thanks{%
E-mail: ariel@albany.edu}}{
address={Department of Physics, University at Albany--SUNY, Albany, NY 12222,USA}}

\begin{abstract}
We use the method of Maximum (relative) Entropy to process information in
the form of observed data and moment constraints. The generic
\textquotedblleft canonical\textquotedblright\ form of the posterior
distribution for the problem of simultaneous updating with data and moments
is obtained. We discuss the general problem of non-commuting constraints,
when they should be processed sequentially and when simultaneously. As an
illustration, the multinomial example of die tosses is solved in detail for
two superficially similar but actually very different problems.
\end{abstract}

\maketitle

\section{Introduction}

The original method of Maximum Entropy, MaxEnt \cite{Jaynes57}, was designed
to assign probabilities on the basis of information in the form of
constraints. It gradually evolved into a more general method, the method of
Maximum relative Entropy (abbreviated ME) \cite{ShoreJohnson80}-\cite%
{Caticha07}, which allows one to update probabilities from arbitrary priors
unlike the original MaxEnt which is restricted to updates from a uniform
background measure.

The realization \cite{CatichaGiffin06} that ME includes not just MaxEnt but
also Bayes' rule as special cases is highly significant. First, it implies
that ME is \emph{capable of reproducing every aspect of orthodox Bayesian
inference} and proves the complete compatibility of Bayesian and entropy
methods. Second, it opens the door to tackling problems that could not be
addressed by either the MaxEnt or orthodox Bayesian methods individually.
The main goal of this paper is to explore this latter possibility: the
problem of processing data plus additional information in the form of
expected values.\footnote{%
For simplicity we will refer to these expected values as \emph{moments}
although they can be considerably more general.}

When using Bayes' rule it is quite common to impose constraints on the prior
distribution. In some cases these constraints are also satisfied by the
posterior distribution, but these are special cases. In general, constraints
imposed on priors do not \textquotedblleft propagate\textquotedblright\ to
the posteriors. Although Bayes' rule can handle \emph{some} constraints, we
seek a procedure capable of enforcing \emph{any} constraint on the posterior
distributions.

After a brief review of how ME processes data and reproduces Bayes' rule, we
derive our main result, the general \textquotedblleft
canonical\textquotedblright\ form of the posterior distribution for the
problem of simultaneous updating with data and moment constraints. The final
result is deceivingly simple: Bayes' rule is modified by a \textquotedblleft
canonical\textquotedblright\ exponential factor. Although this result is
very simple, it should be handled with caution: once we consider several
sources of information such as multiple constraints we must confront the
problem of non-commuting constraints. We discuss the question of whether
they should be processed simultaneously, or sequentially, and in what order.
Our general conclusion is that these different alternatives correspond to
different states of information and accordingly we expect that they will
lead to different inferences.

As an illustration, the multinomial example of die tosses is solved in some
detail for two problems. They appear superficially similar but are in fact
very different. The first die problem requires that the constraints be
processed sequentially. This corresponds to the familiar situation of using
MaxEnt to derive a prior and then using Bayes to process data. The second
die problem, which requires that the constraints be processed
simultaneously, provides a clear example that lies beyond the reach of
Bayes' rule.

\section{Updating with data using the ME method}

Our first concern when using the ME method to update from a prior to a
posterior distribution is to define the space in which the search for the
posterior will be conducted. We wish to infer something about the value of a
quantity $\theta \in \Theta $ on the basis of three pieces of information:
prior information about $\theta $ (the prior), the known relationship
between $x$ \emph{and} $\theta $ (the model), and the observed values of the
data $x\in \mathcal{X}$.\footnote{%
We use the concise notation $\theta $ and $x$ to represent one or many
unknown variables, $\theta =(\theta _{1},\theta _{2}\ldots )$, and one or
multiple experiments, $x=(x_{1},x_{2}\ldots )$.} Since we are concerned with
both $x$ \emph{and} $\theta $, the relevant space is neither $\mathcal{X}$
nor $\Theta $ but the product $\mathcal{X}\times \Theta $ and our attention
must be focused on the joint distribution $P(x,\theta )$. The selected joint
posterior $P_{\text{new}}(x,\theta )$ is that which maximizes the entropy,%
\begin{equation}
S[P,P_{\text{old}}]=-\tint dxd\theta ~P(x,\theta )\log \frac{P(x,\theta )}{%
P_{\text{old}}(x,\theta )}~,~  \label{entropy}
\end{equation}%
subject to the appropriate constraints. All prior information is codified
into the \emph{joint prior} $P_{\text{old}}(x,\theta )=P_{\text{old}}(\theta
)P_{\text{old}}(x|\theta )$. Both $P_{\text{old}}(\theta )$ (the familiar
Bayesian prior distribution) and $P_{\text{old}}(x|\theta )$ (the
likelihood) contain prior information.\footnote{%
The notion that the likelihood function contains prior information may sound
unfamiliar from the point of view of standard Bayesian practice. It should
be clear that the likelihood is \emph{prior} information in the sense that
its functional form is known \emph{before} the actual data is known, or at
least before it can be processed.} The new information is the observed data $%
x^{\prime }$, which in the ME framework must be expressed in the form of a
constraint on the allowed posteriors. The family of posteriors $P(x,\theta )$
that reflects the fact that $x$ is now known to be $x^{\prime }$ is such
that 
\begin{equation}
P(x)=\tint d\theta ~P(x,\theta )=\delta (x-x^{\prime })~.
\label{data constraint}
\end{equation}%
This amounts to an \emph{infinite} number of constraints on $P(x,\theta )$:
for each value of $x$ there is one constraint and one Lagrange multiplier $%
\lambda (x)$.

Maximizing $S$, (\ref{entropy}), subject to the constraints (\ref{data
constraint}) plus normalization, 
\begin{equation}
\delta \left\{ S+\alpha \left[ \tint dxd\theta ~P(x,\theta )-1\right] +\tint
dx\,\lambda (x)\left[ \tint d\theta ~P(x,\theta )-\delta (x-x^{\prime })%
\right] \right\} =0~,
\end{equation}%
yields the joint posterior, 
\begin{equation}
P_{\text{new}}(x,\theta )=P_{\text{old}}(x,\theta )\,\frac{e^{\lambda (x)}}{z%
}~,  \label{solution a}
\end{equation}%
where $z$ is a normalization constant, and $\lambda (x)$ is determined from (%
\ref{data constraint}), 
\begin{equation}
\tint d\theta ~P_{\text{old}}(x,\theta )\frac{\,e^{\lambda (x)}}{z}=P_{\text{%
old}}(x)\frac{\,e^{\lambda (x)}}{z}=\delta (x-x^{\prime })~.
\end{equation}%
The final expression for the joint posterior is%
\begin{equation}
P_{\text{new}}(x,\theta )=\frac{P_{\text{old}}(x,\theta )\,\delta
(x-x^{\prime })}{P_{\text{old}}(x)}=\delta (x-x^{\prime })P_{\text{old}%
}(\theta |x)~,  \label{solution b}
\end{equation}%
and the marginal posterior distribution for $\theta $ is%
\begin{equation}
P_{\text{new}}(\theta )=\tint dxP_{\text{new}}(x,\theta )=P_{\text{old}%
}(\theta |x^{\prime })~,  \label{solution c}
\end{equation}%
which is the familiar Bayes' conditionalization rule.

To summarize: $P_{\text{old}}(x,\theta )=P_{\text{old}}(x)P_{\text{old}%
}(\theta |x)$ is updated to $P_{\text{new}}(x,\theta )=P_{\text{new}}(x)P_{%
\text{new}}(\theta |x)$ with $P_{\text{new}}(x)=\delta (x-x^{\prime })$
fixed by the observed data while $P_{\text{new}}(\theta |x)=P_{\text{old}%
}(\theta |x)$ remains unchanged. We see that in accordance with the minimal
updating philosophy that drives the ME method \emph{one only updates those
aspects of one's beliefs for which corrective new evidence (in this case,
the data) has been supplied}.

\section{Simultaneous updating with moments and data}

Here we generalize the previous section to include additional information
about $\theta $ in the form of a constraint on the expected value of some
function $f(\theta )$, 
\begin{equation}
\tint dxd\theta \,P(x,\theta )f(\theta )=\left\langle f(\theta
)\right\rangle =F~.  \label{<f>}
\end{equation}%
We emphasize that constraints imposed at the level of the prior need not be
satisfied by the posterior. What we do here differs from the standard
Bayesian practice in that we \emph{require} the constraint to be satisfied
by the posterior distribution.

Maximizing the entropy (\ref{entropy}) subject to normalization, the data
constraint (\ref{data constraint}), and the moment constraint (\ref{<f>})
yields the joint posterior,%
\begin{equation}
P_{\text{new}}(x,\theta )=P_{\text{old}}(x,\theta )\frac{e^{\lambda
(x)+\beta f(\theta )}}{z}~,
\end{equation}%
where $z$ is a normalization constant,%
\begin{equation}
z=\tint dxd\theta \,e^{\lambda (x)+\beta f(\theta )}P_{\text{old}}(x,\theta
)\,.
\end{equation}%
The Lagrange multipliers $\lambda (x)$ are determined from the data
constraint, (\ref{data constraint}),%
\begin{equation}
\frac{e^{\lambda (x)}}{z}=\frac{\delta (x-x^{\prime })}{ZP_{\text{old}%
}(x^{\prime })}\quad \text{where}\quad Z(\beta ,x^{\prime })=\tint d\theta
\,e^{\beta f(\theta )}P_{\text{old}}(\theta |x^{\prime })~,
\end{equation}%
so that the joint posterior becomes%
\begin{equation}
P_{\text{new}}(x,\theta )=\delta (x-x^{\prime })P_{\text{old}}(\theta
|x^{\prime })\frac{e^{\beta f(\theta )}}{Z}~.  \label{joint posterior}
\end{equation}%
The remaining Lagrange multiplier $\beta $ is determined by imposing that
the posterior $P_{\text{new}}(x,\theta )$ satisfy (\ref{<f>}). This yields
an implicit equation for $\beta $, 
\begin{equation}
\frac{\partial \log Z}{\partial \beta }=F~.  \label{F}
\end{equation}%
Note that since $Z=Z(\beta ,x^{\prime })$ the resultant $\beta $ will depend
on the observed data $x^{\prime }$. Finally, the new marginal distribution
for $\theta $ is%
\begin{equation}
P_{\text{new}}(\theta )=P_{\text{old}}(\theta |x^{\prime })\frac{e^{\beta
f(\theta )}}{Z}=P_{\text{old}}(\theta )\frac{P_{\text{old}}(x^{\prime
}|\theta )}{P_{\text{old}}(x^{\prime })}\frac{e^{\beta f(\theta )}}{Z}~.
\label{main result}
\end{equation}%
For $\beta =0$ (no moment constraint) we recover Bayes' rule. For $\beta
\neq 0$ Bayes' rule is modified by a \textquotedblleft
canonical\textquotedblright\ exponential factor.

\section{Commuting and non-commuting constraints}

The ME method allows one to process information in the form of constraints.
When we are confronted with several constraints we must be particularly
cautious. In what order should they be processed? Or should they be
processed at the same time? The answer depends on the nature of the
constraints and the question being asked.

We refer to constraints as \emph{commuting} when it makes no difference
whether they are handled simultaneously or sequentially. The most common
example is that of Bayesian updating on the basis of data collected in
multiple experiments: for the purpose of inferring $\theta $ it is
well-known that the order in which the observed data $x^{\prime
}=\{x_{1}^{\prime },x_{2}^{\prime },\ldots \}$ is processed does not matter.
The proof that ME is completely compatible with Bayes' rule implies that
data constraints implemented through $\delta $ functions, as in (\ref{data
constraint}), commute. It is useful to see how this comes about.

When an experiment is repeated it is common to refer to the value of $x$ in
the first experiment and the value of $x$ in the second experiment. This is
a dangerous practice because it obscures the fact that we are actually
talking about \emph{two} separate variables. We do not deal with a single $x$
but with a composite $x=(x_{1},x_{2})$ and the relevant space is $\mathcal{X}%
_{1}\times \mathcal{X}_{2}\times \Theta $. After the first experiment yields
the value $x_{1}^{\prime }$, represented by the constraint $%
c_{1}:P(x_{1})=\delta (x_{1}-x_{1}^{\prime })$, we can perform a second
experiment that yields $x_{2}^{\prime }$ and is represented by a second
constraint $c_{2}:P(x_{2})=\delta (x_{2}-x_{2}^{\prime })$. These
constraints $c_{1}$ and $c_{2}$ commute because they refer to \emph{different%
} variables $x_{1}$ and $x_{2}$. An experiment, once performed and its
outcome observed, cannot be \emph{un-performed} and its result cannot be 
\emph{un-observed} by a second experiment. Thus, imposing one constraint
does not imply a revision of the other.

In general constraints need not commute and when this is the case the order
in which they are processed is critical. For example, suppose the prior is $%
P_{\text{old}}$ and we receive information in the form of a constraint, $%
C_{1}$. To update we maximize the entropy $S[P,P_{\text{old}}]$ subject to $%
C_{1}$ leading to the posterior $P_{1}$ as shown in Figure 1. Next we
receive a second piece of information described by the constraint $C_{2}$.
At this point we can proceed in essentially two different ways:

\noindent \textbf{(a) Sequential updating.} Having processed $C_{1}$, we use 
$P_{1}$ as the current prior and maximize $S[P,P_{1}]$ subject to the new
constraint $C_{2}$. This leads us to the posterior $P_{\text{new}}^{(a)}$.

\noindent \textbf{(b)\ Simultaneous updating.} Use the original prior $P_{%
\text{old}}$ and maximize $S[P,P_{\text{old}}]$ subject to both constraints $%
C_{1}$ and $C_{2}$ simultaneously. This leads to the posterior $P_{\text{new}%
}^{(b)}$.\footnote{%
At first sight it might appear that there exists a third possibility of\
simultaneous updating: (c) use $P_{1}$ as the current prior and maximize $%
S[P,P_{1}]$ subject to both constraints $C_{1}$ and $C_{2}$ simultaneously.
Fortunately, and this is a valuable check for the consistency of the ME
method, it is easy to show that case (c) is equivalent to case (b). Whether
we update from $P_{\text{old}}$ or from $P_{1}$ the selected posterior is $%
P_{\text{new}}^{(b)}$.}%

\begin{figure}[!t]
 \resizebox{.5\columnwidth}{!}
  {\includegraphics[draft=false]{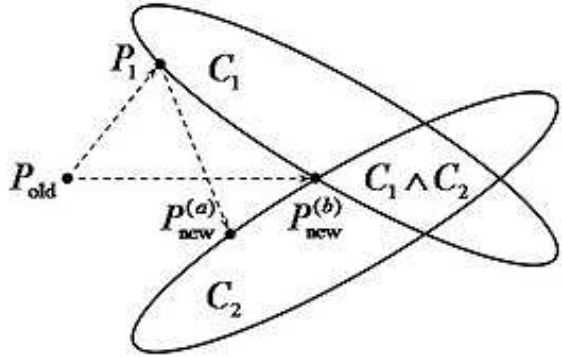}}
  \caption{Illustrating the difference
between processing two constraints $C_{1}$ and $C_{2}$ sequentially ($P_{%
\text{old}}\rightarrow P_{1}\rightarrow P_{\text{new}}^{(a)}$) and
simultaneously ($P_{\text{old}}\rightarrow P_{\text{new}}^{(b)}$ or $P_{%
\text{old}}\rightarrow P_{1}\rightarrow P_{\text{new}}^{(b)}$).}
 \label{Figure1:a}
\end{figure}

To decide which path (a) or (b) is appropriate, we must be clear about how
the ME method treats constraints. The ME machinery interprets a constraint
such as $C_{1}$ in a very mechanical way: all distributions satisfying $%
C_{1} $ are in principle allowed and all distributions violating $C_{1}$ are
ruled out.

Updating to a posterior $P_{1}$ consists precisely in revising those aspects
of the prior $P_{\text{old}}$ that disagree with the new constraint $C_{1}$.
However, there is nothing final about the distribution $P_{1}$. It is just
the best we can do in our current state of knowledge and we fully expect
that future information may require us to revise it further. Indeed, when
new information $C_{2}$ is received we must reconsider whether the original $%
C_{1}$ remains valid or not. Are \emph{all} distributions satisfying the new 
$C_{2}$ really allowed, even those that violate $C_{1}$? If this is the case
then the new $C_{2}$ takes over and we update from $P_{1}$ to $P_{\text{new}%
}^{(a)}$. The constraint $C_{1}$ may still retain some lingering effect on
the posterior $P_{\text{new}}^{(a)}$ through $P_{1},$ but in general $C_{1}$
has now become obsolete.

Alternatively, we may decide that the old constraint $C_{1}$ retains its
validity. The new $C_{2}$ is not meant to revise $C_{1}$ but to provide an
additional refinement of the family of allowed posteriors. In this case the
constraint that correctly reflects the new information is not $C_{2}$ but
the more restrictive $C_{1}\wedge C_{2}$. The two constraints should be
processed simultaneously to arrive at the correct posterior $P_{\text{new}%
}^{(b)}$.

To summarize: sequential updating is appropriate when old constraints become
obsolete and are superseded by new information; simultaneous updating is
appropriate when old constraints remain valid. The two cases refer to
different states of information and therefore \emph{we expect} that they
will result in different inferences. These comments are meant to underscore
the importance of understanding what information is being processed; failure
to do so will lead to errors that do not reflect a shortcoming of the ME
method but rather a misapplication of it.

\section{Sequential updating: a loaded die example}

This is a loaded die example illustrating the appropriateness of sequential
updating. The background information is the following: A certain factory
makes loaded dice. Unfortunately because of poor quality control, the dice
are not identical and it is not known how each die is loaded. It is known,
however, that the dice produced by this factory are such that face $2$ is on
the average twice as likely to come up as face number $5$.

The mathematical representation of this\ situation is as follows. The fact
that we deal with dice is modelled in terms of multinomial distributions.
The probability that casting a $k$-sided die $n$ times yields $m_{i}$
instances for the $i^{th}$ face is%
\begin{equation}
P_{\text{old}}(m|\theta )=P_{\text{old}}(m_{1}...m_{k}|\theta _{1}...\theta
_{k},n)=\frac{n!}{m_{1}!...m_{k}!}\theta _{1}^{m_{1}}...\theta _{k}^{m_{k}}~,
\label{multinomial}
\end{equation}%
where $m=(m_{1},\ldots ,m_{k})$ with $\tsum\nolimits_{i=1}^{k}m_{i}=n$, and $%
\theta =(\theta _{1},\ldots ,\theta _{k})$ with $\tsum\nolimits_{i=1}^{k}%
\theta _{i}=1$. The generic problem is to infer the parameters $\theta $ on
the basis of information about moments of $\theta $ and data $m^{\prime }$.
The additional information about how the dice are loaded is represented by
the constraint $\left\langle \theta _{2}\right\rangle =2\left\langle \theta
_{5}\right\rangle $. Note that this piece of information refers to the
factory as a whole and not to any individual die. The constraint is of the
general form of (\ref{<f>})%
\begin{equation}
C_{1}:\left\langle f(\theta )\right\rangle =F\quad \text{where}\quad
f(\theta )=\tsum\nolimits_{i}^{k}f_{i}\theta _{i}~.
\end{equation}%
For this particular factory $F=0$, and all $f_{i}=0$ except for $f_{2}=1$
and $f_{5}=-2$. Now that the background information has been given, here is
our first example.

We purchase a die. On the basis of our general knowledge of dice we are led
to write down a joint prior 
\begin{equation}
P_{\text{old}}(m,\theta )=P_{\text{old}}(\theta )P_{\text{old}}(m|\theta )~.
\label{P0}
\end{equation}%
(The particular form of $P_{\text{old}}(\theta )$ is not important for our
current purpose so for the sake of definiteness we can choose it flat.) At
this point the only information we have is that we have a die and it came
from a factory described by $C_{1}$. Accordingly, we use ME to update to a
new joint distribution. This is shown as $P_{1}$ in Figure 1. The relevant
entropy is%
\begin{equation}
S[P,P_{\text{old}}]=-\tsum\limits_{m}\tint d\theta ~P(x,\theta )\log \frac{%
P(x,\theta )}{P_{\text{old}}(x,\theta )}~,
\end{equation}%
where 
\begin{equation*}
\tsum\limits_{m}=\tsum\limits_{m_{1}\ldots m_{k}=1}^{n}\delta
(\tsum\nolimits_{i=1}^{k}m_{i}-n)\quad \text{and}\quad \tint d\theta =\tint
d\theta _{1}\ldots d\theta _{k}\,\delta (\tsum\nolimits_{i=1}^{k}\theta
_{i}-1)~,
\end{equation*}%
Maximizing $S$ subject to normalization and $C_{1}$ gives the $P_{1}$
posterior%
\begin{equation}
P_{1}(m,\theta )=\frac{e^{\lambda f(\theta )}}{Z_{1}}P_{\text{old}}(m,\theta
)\,,
\end{equation}%
where the normalization constant $Z_{1}$ and the Lagrange multiplier $%
\lambda $ are determined from%
\begin{equation}
Z_{1}=\tint d\theta \,e^{\lambda f(\theta )}P_{\text{old}}(\theta )\quad 
\text{and}\quad \frac{\partial \log Z_{1}}{\partial \lambda }=F~.  \label{Z1}
\end{equation}%
The joint distribution $P_{1}(m,\theta )=P_{1}(\theta )P_{1}(m|\theta )$ can
be rewritten as%
\begin{equation}
P_{1}(m,\theta )=P_{1}(\theta )P_{\text{old}}(m|\theta )\quad \text{where}%
\quad P_{1}(\theta )=P_{\text{old}}(\theta )\frac{e^{\lambda f(\theta )}}{%
Z_{1}}~.\,  \label{P1}
\end{equation}

To find out more about this particular die we toss it $n$ times and obtain
data $m^{\prime }=(m_{1}^{\prime },\ldots ,m_{k}^{\prime })$ which we
represent as a new constraint%
\begin{equation}
C_{2}:P(m)=\delta (m-m^{\prime })~.  \label{C2a}
\end{equation}%
Our goal is to infer the $\theta $ that apply to our particular die. The
original constraint $C_{1}$ applies to the whole factory while the new
constraint $C_{2}$ refers to the actual die of interest and thus takes
precedence over $C_{1}.$ As $n\rightarrow \infty $ we expect $C_{1}$ to
become less and less relevant. Therefore the two constraints should be
processed sequentially.

Using ME, that is (\ref{solution b}), we impose $C_{2}$ and update from $%
P_{1}(m,\theta )$ to a new joint distribution (shown as $P_{\text{new}%
}^{(a)} $ in Figure 1)%
\begin{equation}
P_{\text{new}}^{(a)}(m,\theta )=\delta (m-m^{\prime })P_{1}(\theta |m)~.
\end{equation}%
Marginalizing over $m$ and using (\ref{P1}) the final posterior for $\theta $
is%
\begin{equation}
P_{\text{new}}^{(a)}(\theta )=P_{1}(\theta |m^{\prime })=P_{1}(\theta )\frac{%
P_{1}(m^{\prime }|\theta )}{P_{1}(m^{\prime })}=\frac{1}{Z_{2}}e^{\lambda
f(\theta )}P_{\text{old}}(\theta )P_{\text{old}}(m^{\prime }|\theta )~.
\label{posterior a}
\end{equation}%
where%
\begin{equation}
Z_{2}=\tint d\theta \,e^{\lambda f(\theta )}P_{\text{old}}(\theta )P_{\text{%
old}}(m^{\prime }|\theta )~.  \label{Z2}
\end{equation}

The readers will undoubtedly recognize that (\ref{posterior a}) is precisely
the result obtained by using MaxEnt to obtain a prior, in this case $%
P_{1}(\theta )$ given in (\ref{P1}), and then using Bayes' theorem to take
the data into account. This familiar result has been derived in some detail
for two reasons: first, to reassure the readers that ME does reproduce the
standard solutions to standard problems and second, to establish a contrast
with the example discussed next.

\section{Simultaneous updating: a loaded die example}

Here is a different problem illustrating the appropriateness of simultaneous
updating. The background information is the same as in the previous example.
The difference is that the factory now hires a quality control engineer who
wants to learn as much as he can about the factory. His initial knowledge is
described by the same prior $P_{\text{old}}(m,\theta )$, (\ref{P0}). After
some inquiries he is told that the only available information is $%
C_{1}:\left\langle \theta _{2}\right\rangle =2\left\langle \theta
_{5}\right\rangle $. Not satisfied with this limited information he decides
to collect data that reflect the production of the whole factory. Randomly
chosen dice are tossed $n$ times yielding data $m^{\prime }=(m_{1}^{\prime
},\ldots ,m_{k}^{\prime })$ which is represented as a constraint,%
\begin{equation}
C_{2}:P(m)=\delta (m-m^{\prime })~.  \label{C2b}
\end{equation}%
The apparent resemblance with (\ref{C2a}) may be misleading: (\ref{C2a})
refers to a single die, while (\ref{C2b}) now refers to the whole factory.
The goal here is to infer the distribution of $\theta $ that describes the
overall population of dice produced by the factory. The new constraint $C_{2}
$ is information in addition to, rather than instead of, the old $C_{1}$:
the two constraints should be processed simultaneously. From (\ref{joint
posterior}) the joint posterior is \footnote{%
As mentioned in the previous footnote, whether we update from $P_{\text{old}}
$ or from $P_{1}$ we obtain the same posterior $P_{\text{new}}^{(b)}$.} 
\begin{equation}
P_{\text{new}}^{(b)}(m,\theta )=\delta (m-m^{\prime })P_{\text{old}}(\theta
|m^{\prime })\frac{e^{\beta f(\theta )}}{Z}~.  \label{joint posterior b}
\end{equation}%
Marginalizing over $m$ the posterior for $\theta $ is 
\begin{equation}
P_{\text{new}}^{(b)}(\theta )=P_{\text{old}}(\theta |m^{\prime })\frac{%
e^{\beta f(\theta )}}{Z}=\frac{1}{\zeta }e^{\beta f(\theta )}P_{\text{old}%
}(\theta )P_{\text{old}}(m^{\prime }|\theta )~,  \label{posterior b}
\end{equation}%
where the new normalization constant is 
\begin{equation}
\zeta =\tint d\theta \,e^{\beta f(\theta )}P_{\text{old}}(\theta )P_{\text{%
old}}(m^{\prime }|\theta )\quad \text{and}\quad \frac{\partial \log \zeta }{%
\partial \beta }=F~.  \label{zeta}
\end{equation}%
This looks like the sequential case, (\ref{posterior a}), but there is a
crucial difference: $\beta \neq \lambda $ and $\zeta \neq Z_{2}$. In the
sequential updating case, the multiplier $\lambda $ is chosen so that the
intermediate $P_{1}$ satisfies $C_{1}$ while the posterior $P_{\text{new}%
}^{(a)}$ only satisfies $C_{2}$. In the simultaneous updating case the
multiplier $\beta $ is chosen so that the posterior $P_{\text{new}}^{(b)}$
satisfies both $C_{1}$ and $C_{2}$ or $C_{1}\wedge C_{2}$. Ultimately, the
two distributions $P_{\text{new}}(\theta )$ are different because they refer
to different problems: $P_{\text{new}}^{(a)}(\theta )$ refers to a single
die, while $P_{\text{new}}^{(b)}(\theta )$ applies to all the dice produced
by the factory.\footnote{%
For the sake of completeness, we note that, because of the peculiarities of $%
\delta $ functions, had the constraints been processed sequentially but in
the opposite order, first the data $C_{2}$, and then the moment $C_{1}$, the
resulting posterior would be the same as for simultaneous update to $P_{%
\text{new}}^{(b)}$.}

\section{Summary and final remarks}

The realization that the ME method incorporates Bayes' rule as a special
case has allowed us to go beyond Bayes' rule to process both data and
expected value constraints simultaneously. To put it bluntly, anything one
can do with Bayes can also be done with ME with the additional ability to
include information that was inaccessible to Bayes alone. This raises
several questions and we have offered a few answers.

First, it is not uncommon to claim that the non-commutability of constraints
represents a \emph{problem} for the ME method. Processing constraints in
different orders might lead to different inferences and this is said to be
unacceptable. We have argued that, on the contrary, the information conveyed
by a particular sequence of constraints is not the same information conveyed
by the same constraints in different order. Since different informational
states should in general lead to different inferences, the way ME handles
non-commuting constraints should not be regarded as a \emph{shortcoming} but
rather as a \emph{feature} of the method.

Second, we are capable of processing both data and moments. Is this kind of
information of purely academic interest or is it something we might
encounter in real life? At this early stage our answer must be tentative: we
have given just one example -- the die factory -- which we think is fairly
realistic. However, we feel that other applications (e.g. in econometrics
and ecology) can be handled in this way as well.\cite%
{GiffinEcon07,GiffinEco07}

Finally, is it really true that this type of problem lies beyond the reach
of Bayesian methods? After all, we can always interpret an expected value as
a sample average in a sufficiently large number of trials. True. We can
always construct a large imaginary ensemble of experiments. Entropy methods
then become in principle \emph{superfluous}; all we need is probability. The
problem with inventing \emph{imaginary} ensembles to do away with entropy in
favor of mere probabilities, or to do away with probabilities in favor of
more intuitive frequencies, is that the ensembles are just what they are
claimed to be, imaginary. They are purely artificial constructions invented
for the purpose of handling incomplete information. It seems to us that a
safer way to proceed is to handle the available information directly as
given (i.e., as expected values) without making additional assumptions about
an imagined reality.

\bigskip
\noindent \textbf{Acknowledgements:} We would like to acknowledge valuable
discussions with C. Cafaro, K. Knuth, and C. Rodr\'{\i}guez.

\section*{Appendix: More on the multinomial problem}

Here we pursue the calculation of the posterior (\ref{posterior b}) in more
detail. To be specific we choose a flat prior, $P_{\text{old}}(\theta )=%
\limfunc{constant}$. Then, dropping the superscript (b),

\begin{equation}
P_{\text{new}}(\theta )=\frac{1}{\zeta _{e}}\delta
(\tsum\limits_{i}^{k}\theta _{i}-1)\tprod\limits_{i=1}^{k}e^{\beta
f_{i}\theta _{i}}\theta _{i}^{m_{i}^{\prime }}.  \label{posterior c}
\end{equation}%
where $\zeta _{e}$ differs from $\zeta $ in (\ref{zeta}) only by a
combinatorial coefficient, 
\begin{equation}
\zeta _{e}=\dint \delta (\tsum\limits_{i}^{k}\theta
_{i}-1)\tprod\limits_{i=1}^{k}d\theta _{i}e^{\beta f_{i}\theta _{i}}\theta
_{i}^{m_{i}^{\prime }}~,  \label{zeta a}
\end{equation}%
and $\beta $ is determined from (\ref{F}) which in terms of $\zeta _{e}$ now
reads $\partial \log \zeta _{e}/\partial \beta =F$. A brute force
calculation gives $\zeta _{e}$ as a nested hypergeometric series, 
\begin{equation}
\zeta _{e}=e^{\beta f_{k}}I_{1}(I_{2}(\ldots (I_{k-1})))\,,
\end{equation}%
where each $I$ is written as a sum of $\Gamma $ functions,

\begin{equation}
I_{j}=\Gamma (b_{j}-a_{j})\dsum\limits_{q_{j}=0}^{\infty }\frac{\Gamma
(a_{j}+q_{j})}{\Gamma (b_{j}+q_{j})~q_{j}!}t_{j}^{q_{j}}I_{j+1}\quad \text{%
with}\quad I_{k}=1~.
\end{equation}%
The index $j$ takes all values from $1$ to $k-1$ and the other symbols are
defined as follows: $t_{j}=\beta \left( f_{k-j}-f_{k}\right) $, $%
a_{j}=m_{k-j}^{\prime }+1$, and 
\begin{equation}
b_{j}=n+j+1+\tsum\limits_{i=0}^{j-1}q_{i}-\tsum\limits_{i=0}^{k-j-1}m_{i}^{%
\prime }~\,,
\end{equation}%
with $q_{0}=m_{0}^{\prime }=0$. The terms that have indices $\leq 0$ are
equal to zero (i.e. $b_{0}=q_{0}=0,$ etc.). A few technical details are
worth mentioning: First, one can have singular points when $t_{j}=0$. In
these cases the sum must be evaluated as the limit as $t_{j}\rightarrow 0.$
Second, since $a_{j}$ and $b_{j}$ are positive integers the gamma functions
involve no singularities. Lastly, the sums converge because $a_{j}>b_{j}$.
The normalization for the first die example, (\ref{Z2}), can be calculated
in a similar way. Currently, for small values of $k$ (less than 10) it is
feasible to evaluate the nested sums numerically; for larger values of $k$
it is best to evaluate the integral for $\zeta _{e}$ using sampling methods.
A more detailed version of the multinomial example is worked out in \cite%
{GiffinEcon07}.

\end{document}